\documentclass[twocolumn,twoside,preprintnumbers,amsmath,amssymb,pacs]{revtex4}
\usepackage{epsfig}
\usepackage{graphicx}
\usepackage{dcolumn}
\usepackage{bm}

\usepackage{pslatex}

 \newcommand\beq{\begin{equation}}
 \newcommand\eeq{\end{equation}}
 \newcommand\beqn{\begin{eqnarray}}
 \newcommand\eeqn{\end{eqnarray}}
 \newcommand{\la}{\langle}  
 \newcommand{\ra}{\rangle}
 \def\mb{\,\mbox{mb}}
\def\fm{\,\mbox{fm}}
\def\GeV{\,\mbox{GeV}}

 \def\Pom{{ I\!\!P}}
 \def\Reg{{ I\!\!R}}

\begin{document}
\hspace*{13cm}
\begin{minipage}{4cm}
{\large USM-TH-183}  \\
\end{minipage}
\bigskip

\title{{\Large Diffraction in QCD}}

\author{Boris Kopeliovich, Irina Potashnikova, Ivan Schmidt}

\affiliation{Departamento de F\'\i sica, Universidad T\'ecnica Federico Santa
Mar\'\i a, Casilla 110-V, Valpara\'\i so, Chile}



\begin{abstract}



This lecture presents a short review of the main features of diffractive
processes and QCD inspired models. It includes the following topics: (1)
Quantum mechanics of diffraction: general properties; (2) Color dipole
description of diffraction; (3) Color transparency; (4) Soft diffraction
in hard reactions:  DIS, Drell-Yan, Higgs production; (5) Why Pomerons
interact weakly; (6) Small gluonic spots in the proton; (7) Diffraction
near the unitarity bound: the Goulianos-Schlein "puzzle"; (8) Diffraction
on nuclei: diffractive Color Glass; (9) CGC and gluon shadowing.
\end{abstract}

\maketitle

\setcounter{page}{1}

\section{Introduction}

Diffraction is associated with the optical analogy, which is elastic 
scattering of light caused by absorption. A new feature of diffraction in 
quantum mechanics is the possibility of inelastic diffraction, which is
nearly elastic scattering with excitation of one or both colliding hadrons
to effective masses which are much smaller that the c.m. energy of 
the collision. The main bulk of diffractive events originate from soft 
interactions. Therefore, it is still a challenge to describe these 
processes starting from the first principles of QCD. Unavoidably, one 
faces the problem of confinement which is still a challenge for the 
theory. Nevertheless, the ideas of QCD help to develop quite an effective 
phenomenology for diffractive reactions, i.e. to establish relations 
between different observables. This lecture presents a mini-review of QCD 
based phenomenological models.

\section{Nonabeliance and diffraction}

Elastic and inelastic diffraction are large rapidity gap (LRG) processes.
Since they emerge as a shadow of inelastic interactions, their amplitudes
are nearly imaginary. This observation is a direct evidence for {\it
nonabeliance} of the underlying theory.

 Indeed, the elastic amplitude can be mediated only by a neutral exchange
in t-channel, therefore the Born graphs in the abelian and nonabelian
cases look like,

 \begin{figure}[htbp]
 \includegraphics[width=5cm]{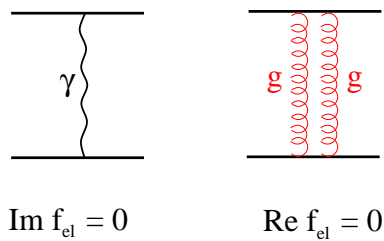}
 \caption{Born approximation for elastic scattering in abelian (left) and 
nonabelian (right) theories.}
 \end{figure}  

The striking difference between these two amplitudes is in their phases.
In the abelian case (e.g. in QED) the Born amplitude is real, while in the 
nonabelian theory (QCD) the amplitude is imaginary.

Data for elastic hadron scattering show that the real part of the
elastic amplitude is small, and this is a direct evidence for
nonabeliance of the underlying dynamics. This is a remarkable
observation, since we have known so far very few manifestations of
nonabeliance in data.

The Born amplitude depicted in Fig.~1 is independent of energy. 
Gluon radiation gives rise to the energy dependence of the total cross
section through the unitarity relation:

 \begin{figure}[htbp]
 \includegraphics[width=8cm]{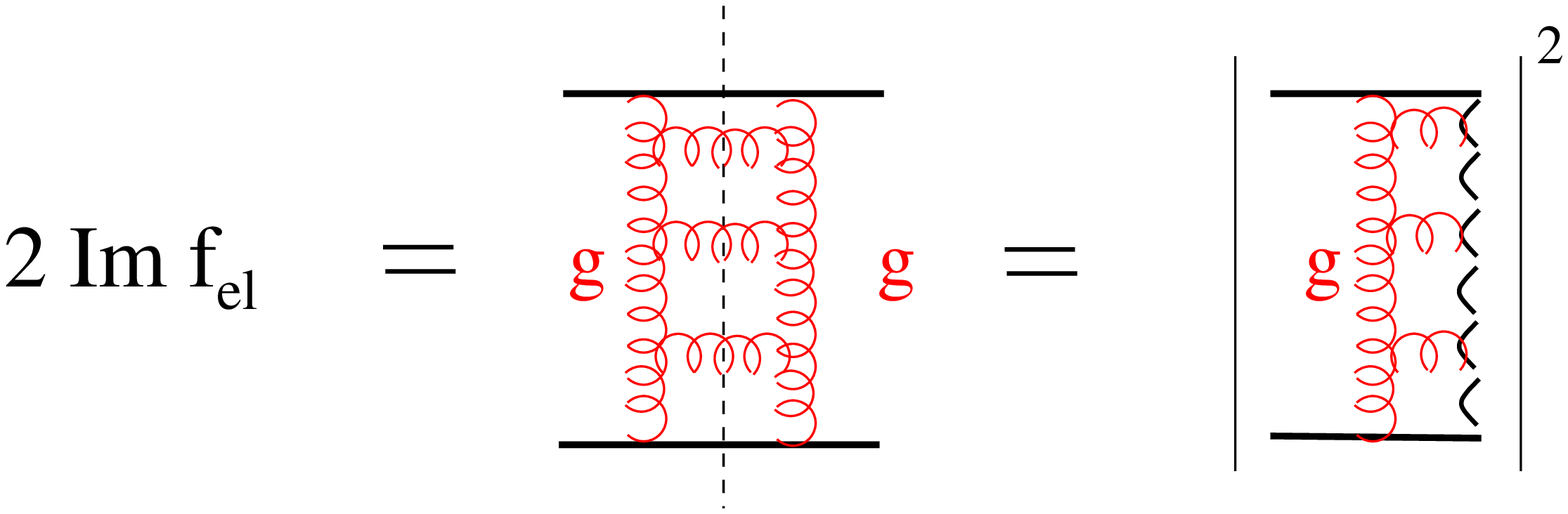}
 \caption{The unitarity relation for the Pomeron amplitude in terms of 
perturbative QCD}
 \end{figure}

 Elastic scattering reaches maximal strength at the unitarity limit of
black disc, ${\rm Im}\, f_{el}(b)=1$,
 \beq
\sigma_{el}=\sigma_{in}=\pi\,R^2
\label{100}
 \eeq
 where $R$ is the radius of interaction.

The unitarity relation tells us that the imaginary part of the partial
amplitude ${\rm Im}\, f_{el}(b)$ cannot rise for ever. After the unitarity
bound is reached, the total cross section can rise only due to an energy
dependence of the interaction radius $R(s)$.  Froissart theorem imposes a
restriction on this, the interaction radius cannot rise with energy faster
than $R\propto \ln(s)$. Then, the total and elastic cross section rise 
with
energy as $\propto \ln^2(s)$ in the Froissart regime of unitarity
saturation.

\section{Regge phenomenology}

In the Regge theory one assumes that the elastic amplitude is mediated by 
exchange of the rightmost singularity in the complex angular momentum 
plane. This singularity is called Pomeron.  

The Regge trajectory corresponding to this singularity is approximately 
linear,
 \beq
\alpha_{\Pom}(t)=\alpha_{\Pom}^0+\alpha_{\Pom}^\prime t
\label{200}
 \eeq
 with parameters 
 \beqn
\alpha_{\Pom}^0 &=& 1.1; \nonumber\\
\alpha_{\Pom}^\prime &=& 0.25\GeV^{-2}
\label{300}
 \eeqn

This behavior follows from data for elastic and total cross sections 
fitted by the formula,
 \beq
f_{el}(t) = \left[i-{\rm
ctg}\frac{\pi\alpha_{\Pom}(t)}{2}\right]\,h(t)\,
\left(\frac{s}{s_0}\right)^{\alpha_{\Pom}(t)}\ ,
\label{400}
 \eeq
 where $h(t)$ is the phenomenological residue function which is not
given by the theory, but is fitted to data. It correlates with the
choice of the parameter $s_0$.

Apparently, the linear $t$-dependence of the Pomeron trajectory
Eq.~(\ref{300}) cannot continue for ever at large negative $t$. Indeed,
the higher order corrections in the ladder graph in Fig.~2 vanish as
powers of the QCD coupling $\alpha_s(t)$ and the Pomeron trajectory
$\alpha_{\Pom}(t)$ should approach the value corresponding to the Born
graph, $\alpha_{\Pom}(t)\to 1$. Indeed, the trajectory seems to level
off at large $|t|$ according to data \cite{brandt}.

It has been a natural and simplest assumption made in the early years of 
the Regge theory that the Pomeron is a Regge pole with a linear 
trajectory and the intercept $\alpha_{\Pom}(t) = 1$. Nowadays, however, we
have a multi-choice answer, and it is still debatable whether the Pomeron 
is

\begin{itemize}

\item
a Regge pole  (probably not, since
$\alpha_{\Pom}^0$
varies with $Q^2$ in DIS);

\item or the DGLAP Pomeron \cite{book}, which corresponds to a specific
ordering for radiated gluons in the ladder graph in Fig.~2, $x\leq
x_{i+1}\leq x_i$ and $k^2_{i+1}<k^2_i\leq Q^2$;

\item or the BFKL Pomeron \cite{bfkl} which does not have ordering in
transverse momenta of radiated gluons, but has no evolution with $Q^2$
either \cite{sardinia};

\item
or something else?

\end{itemize}

\subsection{Triple Regge phenomenology}

The cross section of the single-diffractive process, $a+b \to X +b$ can be 
expressed in terms of the Regge approach. Indeed, if to sum up all final 
state excitations $X$, one can apply the unitarity relation to the 
Pomeron-hadron ($\Pom-a$) amplitude as is shown in Fig.~3.
 \begin{figure}[htbp]
 \includegraphics[width=8cm]{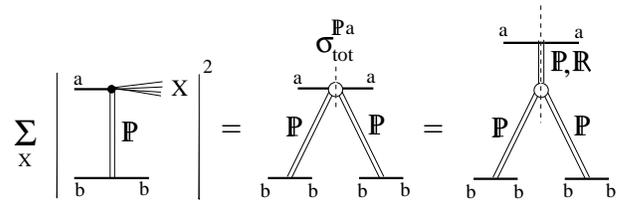}
 \caption{The cross section of single diffraction, $a+b\to X+b$ summed 
over all excitation channels at fixed effective mass $M_X$}
 \end{figure}
 Provided that the effective mass of the excitation is large (but not too
much), $s_0\ll M_X^2\ll s$, one can describe the Pomeron-hadron elastic
amplitude via exchange of the Pomeron or secondary Reggeons in the
$t$-channel. The one arrive to the triple-Regge graph, which corresponds
to the cross section,
 \beq
\frac{d\sigma_{sd}^{ab\to Xb}}{dx_F\,dt} =
\sum\limits_{i=\Pom,\Reg} G_{\Pom\Pom i}(t)
(1-x_F)^{\alpha_i(0)-2\alpha_\Pom(t)}
\left(\frac{s}{s_0}\right)^{\alpha_i(0)-1}
\label{500}
 \eeq
 Here $x_F$ is the Feynman variable for the recoil particle $b$,
$x_F=2p^{||}_b/\sqrt{s} \approx 1-M_X^2/s$.

Equation (\ref{500}) contains new phenomenological functions, effective 
triple-Regge vertices, $G_{\Pom\Pom\Pom}(t)$ and $G_{\Pom\Pom\Reg}(t)$.
The diffractive cross section can also be expressed in terms of 
the Pomeron-hadron total cross section $\sigma^{\Pom a}_{tot}(s'=M_X^2)$.
Most interesting is the asymptotia ($s'=M_X^2\gg s$) of this cross section
related to the triple-Pomeron coupling,
 \beq
G_{3\Pom}(t)=\sigma^{\Pom a}_{tot}\,
N_{\Pom bb}(t)^2\ .
\label{600}
 \eeq
 Here $N_{\Pom bb}(t)$ is the Pomeron - hadron vertex known from $bb$
elastic scattering. Thus, one can extract from data on single diffraction
the Pomeron-hadron total cross section, $\sigma^{\Pom a}_{tot}$, which
carries unique information about the properties of the Pomeron (see 
below).

\section{Quantum mechanics of diffraction}

Diffractive excitation is a nontrivial consequence of presence of
quantum fluctuations in hadrons. In classical mechanics only elastic
scattering is possible. An example is diffractive scattering of
electromagnetic waves.

One can understand the origin of diffractive excitation in terms of
elastic diffraction \cite{fp,gw}. Since a hadron has a composite
structure, different hadronic constituents interact differently causing a
modification of the projectile coherent superposition of states. Such a
modified wave packet is not orthogonal any more to other hadrons different
from the incoming one. This makes possible production of new hadrons, i.e.
diffractive excitation.

To simplify the picture, one can switch to the basis of eigenstates of
interaction. Since a hadron can be excited, it cannot be an eigenstate of
interaction, and can be expanded over the complete set of eigen states 
$|\alpha\ra$ \cite{kl,mp,kst2}:
 \beq
|h\ra = \sum\limits_{\alpha=1}C^h_{\alpha}\,|\alpha\ra\ ,
\label{700}
 \eeq
 which satisfy the condition, $\hat f_{el}|\alpha\ra =
f_\alpha\,|\alpha\ra$, where $\hat f_{el}$ is the elastic amplitude
operator.

Due to completeness and orthogonality of each set of states, the
coefficient $C^h_{\alpha}$ in (\ref{700}) satisfy the relations,
 \beqn
\la h'|h\ra  &=&
\sum\limits_{\alpha=1}(C^{h'}_{\alpha})^*C^h_{\alpha} =
\delta_{hh'}
\nonumber\\
\la \beta|\alpha\ra  &=&
\sum\limits_{h'}(C^{h'}_{\beta})^*C^{h'}_{\alpha} =
\delta_{\alpha\beta}
\label{800}
 \eeqn

The elastic and single diffraction amplitudes can be thus expressed via 
the eigen amplitudes as,
 \beqn
f_{el}^{h\to h} &=& \sum\limits_{\alpha=1}|C^h_{\alpha}|^2\,f_\alpha
\nonumber\\
f_{sd}^{h\to h'} &=&
\sum\limits_{\alpha=1}(C^{h'}_{\alpha})^*C^h_{\alpha}\,f_\alpha
\label{900}
 \eeqn
 Using these expressions and the completeness relations, Eqs.~(\ref{800})
one can calculate the forward single diffraction cross section without
knowledge of the properties of $|h'\ra$,
 \beqn
\left.
\sum\limits_{h'\neq h}\frac{d\sigma^{h\to h'}_{sd}}
{dt}\right|_{t=0} &=&
\frac{1}{4\pi}\left[\sum\limits_{h'}|f_{sd}^{hh'}|^2
-|f_{el}^{hh}|^2\right]\nonumber\\
&=&
\frac{1}{4\pi}\left[\sum\limits_{\alpha}|C^h_{\alpha}|^2\,
|f_\alpha|^2 -\left(\sum\limits_{\alpha}
|C^h_{\alpha}|f_\alpha\right)^2\right]\nonumber\\
&=& \frac{\la f_\alpha^2\ra - \la f_\alpha\ra^2}
{4\pi}
\label{1000}   
 \eeqn
Thus, the forward diffractive cross section is given by the dispersion
of the eigen values distribution. For some specific distributions
the dispersion may be zero. For instance if all the eigen amplitudes are 
equal, or one of them is much larger than others.

According to Eqs.~(\ref{900})-(\ref{1000}) one can calculate the total and
diffractive cross sections on the same footing, provided that the
eigenstates $|\alpha\ra$, their weights $|C^h_{\alpha}|^2$ and the
eigenvalues $f_\alpha$ are known. Notice that the eigen amplitudes
$f_\alpha$ are the same for different hadronic species $|h\ra$. This
remarkable property of eigen amplitudes is employed later on.
 
 In the Froissart regime all the partial eigen amplitudes reach the
unitarity limit, ${\rm Im}\,f_\alpha=1$. Then, according
to the completeness conditions,
 \beqn
f_{el}^{hh}
&\Rightarrow&
\sum\limits_{\alpha=1}|C^h_{\alpha}|^2=1
\nonumber\\
f_{sd}^{hh'} &\Rightarrow&
\sum\limits_{\alpha=1}(C^{h'}_{\alpha})^*C^h_{\alpha}
=0 
\label{1100}
 \eeqn

Diffraction is impossible within a black disc, but only on its periphery,
$b\sim R$. Since in the Froissart regime $R\propto \ln(s)$,
 \beqn
\sigma_{tot}&\propto& \sigma_{el}
\propto \ln^2(s)
\nonumber\\
\sigma_{sd}&\propto& \ln(s)\ ,
\label{1200}
 \eeqn
  i.e. $\sigma_{sd}/\sigma_{tot}\propto 1/\ln(s)$.

 \section{Light-cone color dipole description}

The choice of the eigen state basis depends on the underlying theory.  It
was first realized in \cite{zkl} that the eigenstates of interaction in 
QCD
are color dipoles.  Such dipoles
cannot be excited and can experience only elastic scattering.
Indeed, high energy dipoles have no definite mass, but only separation
$\vec r_T$ which cannot be altered during soft interaction. The
eigenvalues of the total cross section,
$\sigma(r_T)$, also depend on $r_T$, but may also depend on energy.

The total and single diffractive cross sections read,
 \beqn
\sigma_{tot}^{hp} &=&
\sum\limits_{\alpha=1}|C^h_{\alpha}|^2\,\sigma_\alpha
\nonumber\\ &=&
\int d^2r_T\left|\Psi_h(r_T)\right|^2\sigma(r_T) =
\la\sigma(r_T)\ra\ ;
\label{1300}
 \eeqn

 \beqn
&& \left.
\sum\limits_{h'}\frac{d\sigma^{h\to h'}_{sd}}
{dt}\right|_{t=0} =
\sum\limits_{\alpha=1}|C^h_{\alpha}|^2\,
\frac{\sigma_\alpha^2}{16\pi} =
\nonumber\\
&&\int d^2r_T\left|\Psi_h(r_T)
\right|^2\frac{\sigma^2(r_T)}{16\pi} =
\frac{\la\sigma^2(r_T)\ra}{16\pi}
\label{1400}
 \eeqn

The eigenvalue of the cross section for a simplest $\bar qq$ dipole
$\sigma_{\bar qq}(r_T)$ is a fundamental flavor independent quantity. Its
calculation is still a theoretical challenge, but it can be fitted to
data.

A rich source of information about $\sigma_{\bar qq}(r_T)$ is DIS. At
small $x_{Bj}$ the virtual photon exposes hadronic properties as is 
illustrated in Fig.~4.
 \begin{figure}[htbp]
\includegraphics[width=8cm]{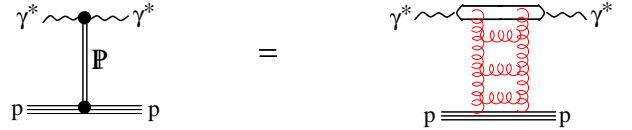}
\caption{The virtual photon interacts via its hadronic fluctuations
which are $\bar qq$ dipoles and more complicated Fock states.
The Pomeron exchange is illustrated as a perturbative ladder.}
 \end{figure}
 One has a control of the dipole size varying the photon virtuality $Q^2$
according to the factorized formula \cite{zkl,nz}
 \beqn
\sigma_{tot}^{\gamma^*p}(Q^2,x_{Bj}) =
\int d^2r_T\int\limits_0^1 dx
\left|\Psi_{\gamma^*}(r_T,Q^2)\right|^2
\sigma_{\bar qq,}(r_T,x_{Bj})
\label{1500}
\eeqn
 One may expect, both intuitively and considering dimensions, that the
mean transverse separation is $\la r_T^2\ra \sim 1/Q^2$. However, the
situation is more complicated than that.

\subsection{The photon distribution amplitudes}

 The dipole size in (\ref{1500}) is governed by the photon $\bar qq$
light-cone wave function $\Psi_{\gamma^*}(r_T,\alpha,Q^2)$ where $\alpha$ 
is the 
fraction of the photon light-cone momentum carried by the quark as is 
illustrated in Fig.~5.
 \begin{figure}[htbp]
\includegraphics[width=3cm]{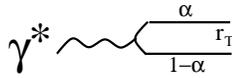}
\caption{Photon virtual dissociation to a $\bar qq$ pair with transverse 
separation $r_T$ and sharing of the light-cone momentum $\alpha$ and 
$1-\alpha$.}
 \end{figure}
 This wave function can be calculated perturbatively \cite{bks}.
 \beqn
\Psi^{T,L}_{\gamma^*}(\vec r_T,\alpha)=
\frac{\sqrt{\alpha_{em}}}{2\,\pi}\,
\bar\chi\,\widehat O^{T,L}\,\chi\,K_0(\epsilon r_T)
\label{1600}
\eeqn
 where $\epsilon^2 = \alpha(1-\alpha)Q^2 + m_q^2$ ;
 \beqn
\widehat O^{T} &=&
m_q\,\vec\sigma\vec e +
 i(1-2\alpha)\,(\vec\sigma\vec n)\,
(\vec {e} \vec\nabla_{r_T})
+ (\vec\sigma\times\vec e)\vec\nabla_{r_T}\ ;
\nonumber\\
\widehat O^{L} &=&
2\,Q\,\alpha(1-\alpha)\,\vec\sigma\vec n
\label{1700}
\eeqn
 It might be confusing that these wave functions are not normalized, the 
transverse part is even divergent. Therefore it is better to call them 
distribution amplitudes.

 The mean transverse $\bar qq$ separation for a transversely polarized 
photon is,
 \beqn
\la r_T^2\ra \sim \frac{1}{\epsilon^2}=
\frac{1}{Q^2 \alpha(1-\alpha)+m_q^2}\ ,
\label{1800} 
\eeqn
 i.e. the separation is about as small as $1/Q^2$, except the endpoints
$\alpha\to 0$ or $1$. Notice that $m_q\sim \Lambda_{QCD}$ plays here role 
of an
infra-red cut off.

\subsection{Dipole cross section and color transparency}

 The central ingredient of Eq.~(\ref{1500}) is the phenomenological
universal cross section $\sigma_{\bar qq}(r_T,x)$ for the interaction of a
nucleon with a $\bar qq$ dipole of transverse separation $r_T$. It must me
energy dependent due to teh higher-order corrections shown in Fig.~4. In
the presence of hard scale the dimensioneless quantity must be $s/Q^2=1/x$
where $x$ is the Bjorken variable. The parametrization suggested in
\cite{gbw},
 \beq
\sigma_{\bar qq}(r_T,x)=\sigma_0\,\left[
1-e^{-{1\over4}\,r_T^2\,Q_s^2(x)}\right]\ ,
\label{1900}
 \eeq
 successfully fits HERA data for the proton structure function
$F_2(x,Q^2)$ at small $x$ with parameters: $Q_s(x)=1\GeV \times
(x_0/x)^{\lambda/2}$ and $\sigma_0=23.03\mb$; $\lambda=0.288$;
$x_0=3.04\cdot 10^{-4}$. This cross section incorporates the phenomenon of
saturation at a soft scale, since it levels off at large separations,
$r_T^2\gg 1/Q_s^2$.

A remarkable feature of this dipole cross section is Color Transparency
(CT), namely for small dipoles, $r_T\to 0$, the cross section vanishes as
$\sigma_{\bar qq}(r_T)\propto r_T^2$ \cite{zkl}. This is a much more
general property of any dipole cross section in QCD, since a point-like
colorless object cannot interact with external color fields. The quadratic
$r_T$-dependence is a direct consequence of gauge invariance and
nonabeliance of QCD.

The effect of CT has been searched for in different reactions. In some of
them, quasielastic high-$p_T$ scattering of electrons \cite{slac} and
hadrons \cite{bnl}, no unambiguous signal of CT was observed. Those
processes turned out to be unsuitable for CT searches \cite{jennings,jan},
since the formation length of the hadrons was too short compared to the
nuclear size.

More successful was search for CT in diffractive leptoproduction of vector 
mesons, proposed in \cite{knnz} and confirmed by the E665 experiment
\cite{e665}. This process illustrated in Fig.~6 is different from DIS by 
projection of the produced $\bar qq$ dipole on the vector meson wave 
function. 
 \begin{figure}[htbp]
\includegraphics[width=7cm]{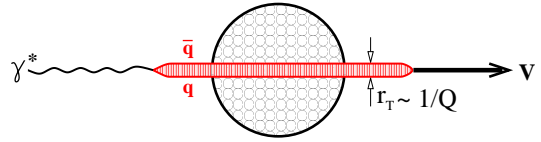}
\caption{Quasielasic virtual photoproduction of vector mesons.
At high $Q^2$ the $\bar qq$ dipole experiences little attenuation in the 
nucleus.}
 \end{figure}
 This projection suppresses the endpoint part of the distribution
amplitude and makes the signal of CT stronger. A new measurement of the
effect of CT was done recently by the HERMES experiment \cite{hermes}.
They found a good signal of CT in accordance with theoretical predictions
\cite{knst-ct}.

Another diffractive process suggested in \cite{bk}, coherent production of
high-$p_T$ back-to-back jets on nuclei, also revealed a strong signal of
CT \cite{e791} in good agreement with theoretical estimates \cite{fms}. In
this process the nucleus remains intact, which is possible due to
sufficiently high energy. A bright signature of CT observed in the E791
experiment at Fermilab is the $A$-dependence of the coherent cross
section, $\propto A^{4/3}$. This corresponds to full nuclear transparency.

\section{Soft diffraction in hard reactions}

\subsection{Diffractive DIS}

The contribution of diffractive quasielastic production of vector mesons
(see Fig.~6) is a tiny fraction, vanishing as $1/Q^2$, of the total
inclusive DIS cross section. However the fraction of all diffractive
events associated with large rapidity gaps in DIS is large, about $10\%$,
and is nearly independent of $Q^2$. It turns out to be a result of a
contribution of rare soft fluctuations in the hard photon. According to
(\ref{1800}) a longitudinally asymmetric $\bar qq$ pair with $\alpha$ or
$1-\alpha\sim 1/Q^2$ have a large hadronic size and experience soft
diffractive interactions like hadrons. Although the admixture of such soft
fluctuations in the virtual photon is tiny, that may be compensated by a
large interaction cross section. This interplay between the fluctuation
probability and the cross section is illustrated for inclusive and
diffractive DIS in Table.~1 \cite{kp-soft}.
 \begin{table}[htbp]
\caption{Interplay between the probabilities of hard and soft fluctuations 
in a highly virtual photon and the cross section of interaction of these 
fluctuations.}\vspace*{0.3cm}
\begin{tabular}{|c|c|c|c|c|}
 \hline
 \vphantom{\bigg\vert}
    &
$|C_\alpha|^2$
  & $\sigma_\alpha$
  &
$\sigma_{tot}\!=\!\!\!\!\!\sum\limits_{\alpha=soft}^{hard}|C_\alpha|^2
\sigma_\alpha$
  &
$\sigma_{sd}\!=\!\!\!\!\!\sum\limits_{\alpha=soft}^{hard}|C_\alpha|^2
\sigma^2_\alpha$
   \\
[3mm]
\hline &&&&\\
Hard& $\sim 1$ & $\sim\frac{1}{Q^2}$ &
$\sim \frac{1}{Q^2}$
& $\sim \frac{1}{Q^4}$  \\
[3mm]
\hline &&&&\\
Soft & $\sim \frac{m_q^2}{Q^2}$ &
$\sim\frac{1}{m_q^2}$ &
$\sim\frac{1}{Q^2}$ &
$\sim\frac{1}{m_q^2Q^2}$
  \\
[3mm]
\hline
\end{tabular}
 \end{table}
 Hard fluctuations of the photon have large weight, but vanishing as
$1/Q^2$ cross section, while soft fluctuations have a small, $m_q^2/Q^2$,
weight, but interact strongly, $\sigma\sim 1/m_q^2$. The latter factor
compensates the smallness of the probability in the case of DIS, and
over-compensates it for diffraction.

 Thus, we conclude that inclusive DIS is semi-hard and semi-soft, and the
soft component is present at any high $Q^2$. On the other hand,
diffractive DIS (called sometimes "hard diffraction") is predominantly a
soft process. This is why its fraction in the total DIS cross section is
nearly $Q^2$-independent. One can test this picture studying the $Q^2$ 
dependence of the diffractive DIS \cite{beatriz1}.

Since diffraction is a source of nuclear shadowing \cite{gribov},
that also should scale in $x$. Indeed, most of experiment have not 
found any variation with $Q^2$ of shadowing in DIS on nuclei. Only the NMC 
experiment managed to find a weak scaling violation which agrees with
theoretical expectations \cite{krt}.

Notice that in spite of independence of $Q^2$, both diffraction and 
shadowing are higher twist effects. This is easy to check considering 
photoproduction of heavy flavors. In this case the hard scale is imposed 
by the heavy quark mass, and diffraction becomes a hard process with cross 
section vanishing as $1/m_Q^4$. Nuclear shadowing also vanishes as 
$1/m_Q^2$.

The true leading twist diffraction and shadowing are associated with gluon 
radiation considered below. 

\subsection{Diffractive Drell-Yan reaction} 

The dipole description of the Drell-Yan reaction in many respects 
is similar to DIS. This is not a surprize, since the two processes are 
related by QCD factorization. The cross section of heavy photon
($\gamma^*\to \bar ll$) radiation by a quark reads \cite{k,kst1,krt3,bhq},
 \beq
\frac{d\sigma(qp\to \gamma^*X)}{d\ln\alpha}
=\int d^2r_T\, |\Psi^{T,L}_{\gamma^* q}(\alpha,r_T)|^2
    \sigma_{q\bar q}(\alpha r_T,x),
\label{1950}
 \eeq
 Here $\alpha$ is the fraction of the quark light-cone momentum taken away
by the dilepton; $r_T$ is the photon-quark transverse separation; and the
light-cone distribution function $\Psi$ is similar to one in DIS,
Eq.~(\ref{1600}), and can be found in \cite{k,kst1,krt3}.

Notice that the dileptons are radiated only in the fragmentation region of 
the quark and are suppressed at mid rapidities. Indeed, due to CT the 
dipole cross section vanishes as $\sigma_{q\bar q}(\alpha r_T,x)\propto 
\alpha^2$ at $\alpha\to 0$.

There is an important difference between DIS and DY reaction. In the
inclusive DIS cross section one integrates over $0<\alpha<1$, this is why
this cross section is always a mixture of soft and hard contributions (see
Table~1). In the case of DY reaction there is a new variable, $x_1$, which
is fraction of the proton momentum carried by the dilepton. Since $\alpha
> x_1$, one can enhance the soft part of the DY cross section selecting
events with $x_1\to 1$. This soft part of the DY process is subject to 
unitarity corrections \cite{beatriz3} which are more important than in 
DIS \cite{beatriz4}.

Another distinction between DIS and DY is suppression of the DY
diffractive cross section. Namely, the forward cross section of
diffractive radiation $qp\to \bar llqp$ is zero \cite{kst1}. Indeed,
according to (\ref{1000}) the forward diffractive cross section is given
by the dispersion of the eigen amplitude distribution. However, in both
eigen states $|q\ra$ and $|q\gamma^*\ra$ only quark interacts. So the two
eigen amplitudes are equal, and the dispersion is zero.

Nevertheless, in the case of hadronic collision diffractive DY cross
section does not vanish in the forward direction. In this case the two
eigen states are $|\bar qq\ra$ and $|\bar qq\gamma^*\ra$ (for the sake of
simplicity we take a pion). The interacting component of these Fock states
is the $\bar qq$ dipole, however it gets a different size after the $q$ or
$\bar q$ radiate the photon. Then the two Fock states interact
differently, and this leads to a nonvanishing forward diffraction. Notice
that the diffractive cross section is proportional to the dipole size
\cite{ks-t}.

\subsection{Diffractive Higgs production}

Diffractive higgsstrahlung is rather similar to diffractive DY, since in
both cases the radiated particle does not take part in the interaction
\cite{ks-t}. However, the Higgs coupling to a quark is proportional to the
quark mass, therefore, the cross section of higgsstrahlung by light
hadrons is vanishingly small.

A larger cross section may emerge due to admixture of heavy flavors in
ligt hadrons. A novel mechanism of exclusive Higgs production, $pp\to H
pp$, due to direct coaliscence of heavy quarks, $\bar QQ\to H$ was
proposed in \cite{bkss}. The cross section of Higgs production was
evaluated ssuming $1\%$ of intrinsic charm (IC) \cite{stan} and that
heavier flavors scale as $1/m_Q^2$ \cite{maxim}. The results are shown in
Fig.~7 as function of Higgs mass for different intrinsic heavy flavors.
 \begin{figure}[htbp]
 \includegraphics[width=7cm]{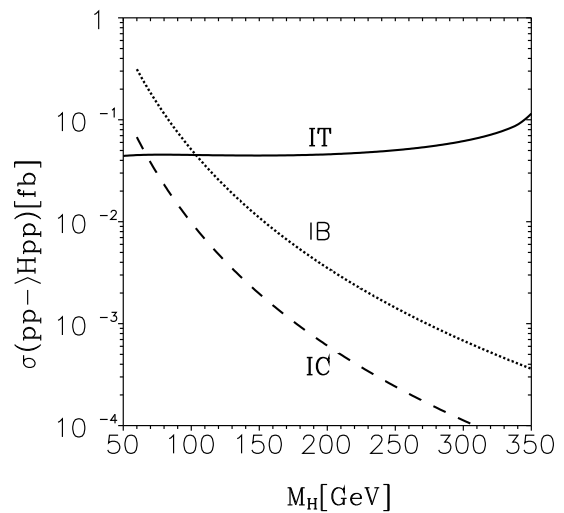}
 \caption{Cross section of exclusive diffractive Higgs production, $pp\to 
Hpp$, from intrinsic charm (IC), bottom (IB) and top (IT) \cite{bkss}.}
 \end{figure}

\section{Diffractive Excitation of Hadrons}

\subsection{Excitation of the valence quark skeleton}

A hadron can be excited in diffractive reaction $hp\to Xp$ by different
mechanisms. One possibility is to excite the valence quark skeleton
without gluon radiation \cite{mine,kps}. This process is illustrated in
Fig.~8
 \begin{figure}[htbp]
 \includegraphics[width=8cm]{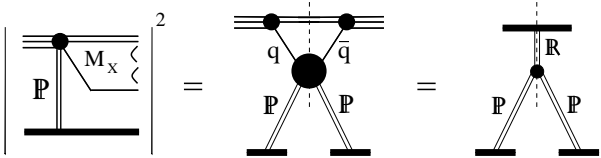}
 \caption{Diffractive excitation of the valence quark skeleton of a 
hadron.}
 \end{figure}
 A clear signature of this process is the dependence of the cross section
on effective mass of the excitation, $M_X$. It must be
$d\sigma_{sd}/dM_X^2\propto 1/M_X^3$, since is related to the intercept of
the secondary Reggeon, $\alpha_\Reg(0)=1/2$, as is demonstrated in Fig.~8.

The specific mass-dependence allows to single out this contribution from
data for diffractive reaction $pp\to pX$, Using the results of the
triple-Regge analysis of data \cite{kklp} we can evaluate the relative
probability of excitation with no gluon radiation \cite{kps},
 \beqn
R_{sd}=\left.\frac{d\sigma_{sd}/dp_T^2}
{d\sigma_{el}/dp_T^2}\right|_{p_T=0} =
\frac{5.5\mb/\GeV^2}{84.5\mb/\GeV^2}=0.065
\label{2000}
\eeqn
 This fraction turns out to be very small, only few percent from the
forward elastic cross section. This suppression can be understood as
follows. In terms of duality the triple-Regge graph in Fig.~8 is
equivalent to excitation of nucleon resonances. Their and proton wave
functions are orthogonal. Therefore the matrix element
$\la\Psi_p(r_T)|\sigma(r_T)|\Psi_X(r_T)\ra$ is not zero only due to
variation of the dipole cross section with $r_T$. However, since the
dipole cross section levels off at large separations, only the short range
part of the integration, $r_T<1/Q_s\sim 0.3\fm$ contributes to the overlap
integral. This explains the observed smallness of gluonless diffraction
\cite{kps}.

\subsection{Diffractive gluon radiation}

A hadron can be excited differently, by shaking off a part of its gluonic
field in the form of gluon radiation. Since gluons are vector particles,
they can propagate through large rapidity intervals without attenuation
and carry a tiny fraction of the hadron momentum. Therefore the effective
mass of the excitation should be large. This is a decisive signature of
radiation, the high-mass tail of the diffractive cross section,
$d\sigma_{sd}/dM_X^2\propto 1/M_X^2$.  Observation of this behavior is
undebatable proof for excitation via gluon radiation, and the cross
section can reliably singled out of data \cite{kklp}.

Fig.~9 shows how the cross section of diffractive gluon radiation is
related to the triple-Pomeron graph. According to Eq.~(\ref{600}) it can
also be expressed in terms of the Pomeron-proton total cross section
$\sigma^{\Pom p}_{tot}(M_X^2)$, for which the effective mass of the
excitation plays role of the c.m. energy.
 \begin{figure}[htbp]
\includegraphics[width=8cm]{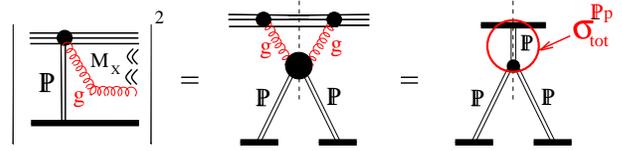}
\caption{The cross section of diffractive gluon radiation related to the 
triple-Pomeron term, or to the Pomeron-proton total cross section.}
 \end{figure}

The triple-Regge fit to data \cite{kklp} with parametrization (\ref{500})
reliably fixes the triple-Pomeron term which provides unique information
about $\sigma^{\Pom p}_{tot}$. Since the Pomeron is a gluonic colorless
dipole, one should probably expect a cross section about $9/4$ times
larger than for mesons. Therefore, one expects $\sigma^{\Pom p}_{tot}\sim
50\mb$.

Surprisingly, data analyses \cite{kaidalov,dino,schlein} lead to
 \beqn
\sigma^{\Pom p}_{tot}\sim 2\mb\ (!)
\label{2100}
\eeqn

\subsection{Small gluonic spots}

Why does the Pomeron interact so weakly, while gluons interact stronger 
than quarks?

One should recall color transparency: the Pomeron-proton cross section
vanishes if the transverse size of the Pomeron (gluonic dipole) is small.

This effect of smallness of gluonic dipoles cannot be explained in pQCD
which treats quarks and gluons as free particles. This is a
nonperturbative phenomenon which may be related to the small size of
gluonic fluctuations in the instanton-liquid model \cite{shuryak}. It is
also supported by calculations on the lattice \cite{pisa} which reveal a
very short gluon correlation radius.

The shape of the impact-parameter distribution of gluons is not known, but
important is the mean size of the dipole. This size $r_0$ treated as a
phenomenological parameter was fixed at $r_0=0.3\fm$ by a fit to
diffraction data \cite{kst2}. Thus, we arrive at an image of the proton
shown in Fig.~10.
 \begin{figure}[htbp]
\includegraphics[width=2cm]{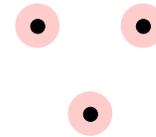}
 \caption{Light-cone snap-shot of the
proton.}
 \end{figure}
 Gluons in the proton are located in small spots of size $0.3\fm$. We
employ this picture in what follows and provide more evidence for it.

\section{Total and Elastic Cross Sections}

Presence of the semihard scale $r_0$ allows to use pQCD
to calculate in a parameter-free way the cross section of gluon
bremsstrahlung rising with energy.
The calculations performed in \cite{k3p} confirm this. 

\subsection{Total cross section}

The hadronic cross section was found in \cite{k3p,calabria} to have the 
following structure,
 \beq
\sigma_{tot}=\sigma_0 + \sigma_1\,
\left(\frac{s}{s_0}\right)^\Delta\ .
\label{2200}
 \eeq
 Here $\sigma_0$ is the term related to hadronic collisions without gluon
radiation. In the string model, for instance, it corresponds to string
crossing and flipping. This part of the cross section is independent of
energy, since is related to the Lorentz invariant transverse size of the
quark skeleton.

 The second term in (\ref{200})  is related to the contribution of gluon
bremsstrahlung to the total cross section. Since this part is expected to
be as small as $r_0^2$, $\sigma_1$ should be small either. Indeed, it was
found in \cite{k3p} that $\sigma_1=27/4\,C\,r_0^2$, where factor $C\approx
2.4$ is related to the behavior at small separations of the dipole-proton
cross section calculated in Born approximation, $\sigma(r_T)= Cr_T^2$ at
$r_T\to0$.

The energy dependence of the second term in (\ref{2200}) was found to be
rather steep, 
 \beqn
\Delta=\frac{4\alpha_s}{3\pi}\approx 0.17
\label{2300}
 \eeqn
 This exponent seems to be too large compared to the experimentally
measured $\sigma_{tot}\propto s^\epsilon$ with $\epsilon\approx 0.1$.
Nevertheless, formula (\ref{2200}) describe data well as is shown in
Fig.~11.
 \begin{figure}[htbp]
\includegraphics[width=7cm]{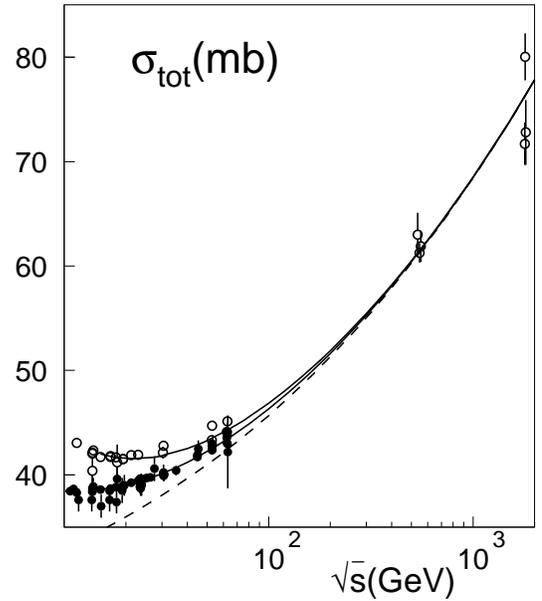}
\caption{Total $pp$ (closed points) and $\bar pp$ (open points) cross 
sections. Dashed curve is the Pomeron contribution Eq.~(\ref{2200}) with 
one parameter $\sigma_0$ adjusted to a single experimental point at 
$\sqrt{s}=540\GeV$, other parameters are calculated. Solid curves include
Reggeons which are fitted to data to describe the cross section at low 
energies.}
 \end{figure}
 This is not a surprise, the energy dependence is slowed down by presence
of the first energy independent term in (\ref{2200}). If to approximate
the cross section (\ref{2200}) by the simple power dependence
$s^\epsilon$, then the effective exponent reads,
 \beq
\epsilon=\frac{\Delta}{1+\sigma_0/\sigma_1\,(s/s_0)^{-\Delta}}
\label{2400}
 \eeq
 So, one should expect the steepness of the energy dependence of
the total cross section to rise with energy. 

\subsection{Elastic slope}

The mean size, $\la r^2(s)\ra$, of the gluonic spots (Fig.~10) rises with
energy due to Brownian motion performed by multiply radiated gluons.
The speed of this growth is related to the slope parameter, 
$\alpha_\Pom^\prime$ of the Pomeron trajectory,
 \beqn
{1\over4}\,
\frac{d\la r^2(s)\ra}{d\ln(s)}=
\alpha^\prime_{\Pom}= 
\frac{\alpha_s}{3\pi}\,r_0^2
=  0.1\GeV^{-2}
\label{2500}
 \eeqn

In the Regge approach this phenomenon is related to the $t$-slope of the
differential elastic cross section,
$B_{el}(s)=B_0+2\alpha^\prime_{\Pom}\ln(s)$.

Using (\ref{2500}) and data for electromagnetic proton formfactor to
calculate $B_0$, the energy dependent elastic slope was calculated in
\cite{k3p} in good agreement with data shown in Fig.~12.
 \begin{figure}[htbp]
\includegraphics[width=7cm]{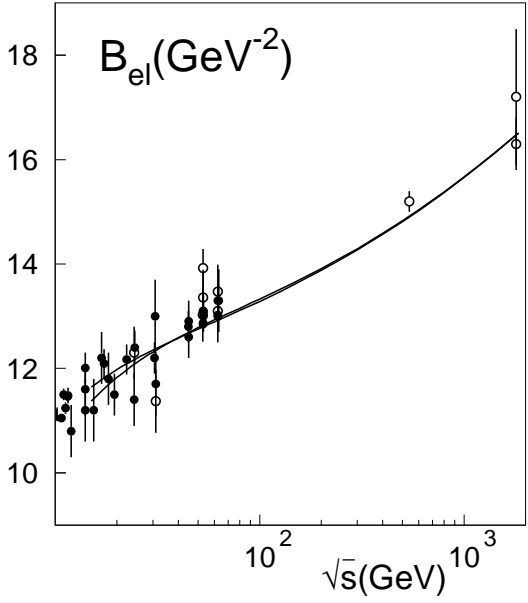}
 \caption{Slope parameter for $pp$ (closed points) and $\bar pp$ (open
points) elastic cross sections.  Solid curves include Reggeons which fitted
to data for total cross section (Fig.~11).}
 \end{figure}
 Notice that the amplitude Eq.~(\ref{2200}) was unitarized in these 
calculations what is important for $pp$ scattering (see next section).

\section{Saturation of the unitarity bound}

The mean number of radiated gluons slowly rises with energy, as well as
the mean size of the gluonic clouds in Fig.~10. This gives rise to an
energy dependence of the interaction radius in accordance with
(\ref{2500}).

However, the predicted value (\ref{2500}) of parameter
$\alpha^\prime_\Pom$ is rather small compared to the effective one, known
from $pp$ data, $\alpha^\prime_{eff}=0.25\GeV^{-2}$. Why does the slope
shown in Fig.~12 rise so steeply with energy? Does it contradict
theoretical expectations?

\subsection{Onset of the Froissart regime}

There is another source of energy dependence of the interaction radius
related to the clotheness of the unitarity bound. The rise of the total
cross section can originate either from the rise of the partial amplitude,
or from increase of the interaction radius. In the vicinity of the
unitarity bound, ${\rm Im}\,f_{el}(b)\leq 1$, the partial amplitude cannot
rise any more in central collisions, while on the periphery the amplitude
is small and there is still room for growth. This is how the interaction
radius rises \cite{k3p,calabria}.

In the Froissart regime (unitarity saturation) the interaction area rises
$\propto \ln^2(s)$ resulting in a fast shrinkage of the diffraction cone.
In the regime of saturation $\alpha^\prime_{eff}\gg \alpha^\prime_{\Pom}$.
Apparently, an onset of this phenomenon explains why the elastic slope,
$B_{el}={1\over2}\la b^2\ra$, rises with energy much steeper than is
predicted by (\ref{2500}).

In order to figure out whether the unitarity limit is indeed reached in
$pp$ collisions at high energies, one can explicitly check with the
partial amplitude related via Fourier transformation to the measured
elastic differential cross section. This is depicted in Fig.~13.
 \begin{figure}[htbp] \includegraphics[width=8cm]{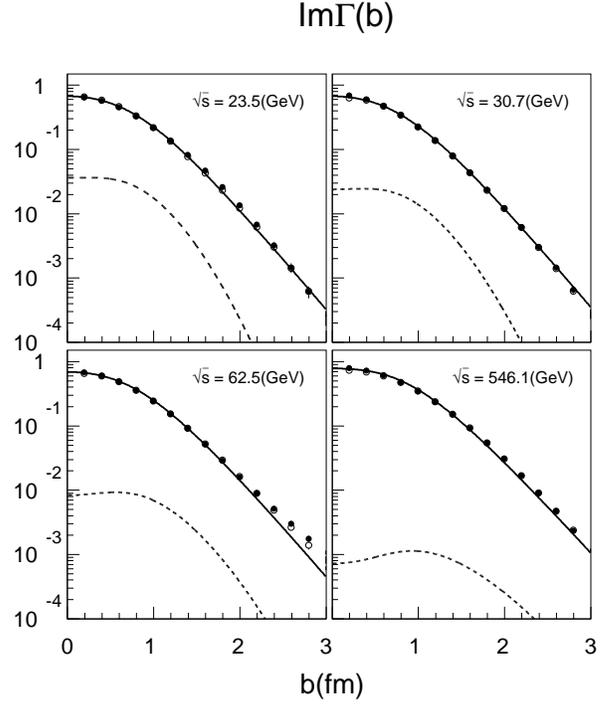}
\caption{Imaginary part of the partial elastic amplitude as function of
impact parameter. Points are the Fourier transformed experimental data for
differential cross section \cite{k3p}. Solid curves show the results of
calculation with the unitarized amplitude Eq.~(\ref{2200}). Dashed curves
show the contribution of the secondary Reggeons strongly shadowed by the
unitarity corrections (see \cite{k3p} for details). }
 \end{figure}
Indeed, both the data and theory demonstrate a nearly-saturation of the
unitarity bound for central $pp$ collisions.

\subsection{Far from saturation: \boldmath$J/\Psi$ production}

In order to test whether a substantial part of the observed
$\alpha^\prime_{eff}$ indeed originates from saturation, one should
look at a diffractive reaction for which the partial amplitude is far
from the unitarity bound. Then one should expect the parameter
$\alpha^\prime_{eff}$ to get its genuine value $\alpha^\prime_{eff}=
\alpha^\prime_\Pom=0.1\GeV^{-2}$ without absorptive (unitarity) 
corrections \cite{qm02}.

$J/\Psi$-proton elastic scattering would be very suitable for this 
purpose. Indeed, the partial amplitude for central collision (b=0) can be 
evaluated as,
 \beq
f_{el}^{\Psi p}(0) = \frac{\sigma^{\Psi p}_{tot}}
{4\pi\,B^{\Psi p}_{el}} = 0.3\ .
\label{2600}
 \eeq
 For this estimate we assumed the energy range of HERA $\sqrt{s}\sim 
100\GeV$ and used $\sigma^{\Psi p}_{tot}=6\mb$ \cite{hikt} and $B^{\Psi 
p}_{el}=4\GeV^{-2}$ \cite{zeus1}.

Thus, the partial amplitude for $J/\Psi$-proton elastic scattering is
safely quite below the unitarity bound at all impact parameters. Then one
should observe a reduced value of $\alpha^\prime_{eff}$ in accordance with
(\ref{2500}).

Of course $J/\Psi$-proton scattering is not accessible, but it can be 
replaced by elastic photoproduction, $\gamma p\to J/\Psi p$, having in 
mind vector dominance. Although vector dominance is a poor approximation 
for charmonia \cite{vdm}, this only enforces our statements, since the 
real $\bar cc$ dipole is smaller than $J/\Psi$.

The measurements of energy dependent $t$-slope of the photoproduction cross 
section was performed by ZEUS collaboration at HERA. The result for the 
Pomeron trajectory are shown in Fig.~14.  
 \begin{figure}[htbp]
\includegraphics[width=8cm]{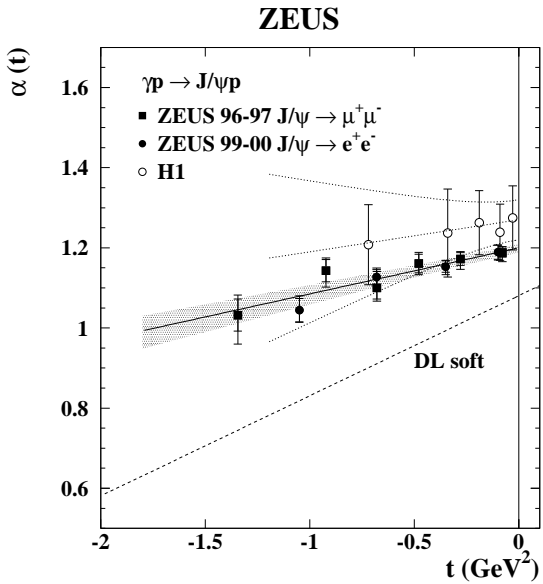}
\caption{The Pomeron trajectory $\alpha_\Pom(t)$ for elastic 
photoproduction of $J/\Psi$ measured at $\sqrt{s}=35-350\GeV$ 
\cite{zeus1}.
the solid line is a fit (see text).}
 \end{figure}
 These data fitted with $\alpha_\Pom(t) = \alpha^0_\Pom +
\alpha^\prime_\Pom\,t$ result in,
 \beqn
\alpha^0_\Pom &=& 0.2 \pm 0.009;\nonumber\\
\alpha^\prime_\Pom &=& 0.115 \pm 0.018\GeV^{-2}\ .
\label{2700}
 \eeqn
 The value of $\alpha^\prime_\Pom$ is in good agreement with the 
prediction (\ref{2500}) made in \cite{k3p}.

\subsection{Goulianos-Schlein puzzle}

The unitarity or absorptive corrections are especially significant for
off-diagonal diffractive channels. As was demonstrated in (\ref{1100})  
diffraction is completely terminated in the unitarity limit. As far as
we already have an onset of the Froissart regime, diffraction has to be
suppressed, and the higher the energy is, the more.

The absorptive corrections to the diffractive amplitude have the form
 \beqn
f_{sd}(b) \Rightarrow
f_{sd}(b)\,\left[1-
{\rm Im}\,f_{el}(b)\right]
\label{2800}
 \eeqn
 Thus, at the unitarity bound,
${\rm Im}\,f_{el}(b)\to 1$, diffraction vanishes everywhere except the 
very periphery. 

Do we see any suppression in data?\\ {\it Yes}, a strong deviation from
the nonunitarized Regge model was found in \cite{dino,schlein} for single
diffraction. Comparison of available data with the triple-Regge prediction
uncorrected for unitarity, is shown in Fig.~15.
 \begin{figure}[htbp]
\includegraphics[width=8cm]{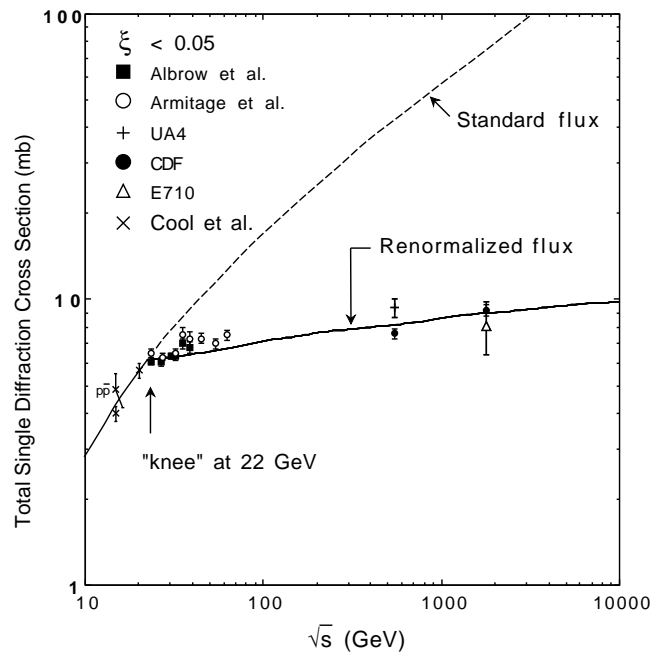}
\caption{Data for single diffraction in comparison with extrapolation of 
the nonunitarized Regge model (dashed curve) \cite{dino}}
 \end{figure}

 Notice that the effect of unitarity corrections Eq.~(\ref{1800}) cannot
be reproduced by a simple suppression of the Pomeron flux. The flux
damping or renormalization factor is independent of impact parameter,
while this dependence is the central issue in (\ref{1800}) (see in 
\cite{beatriz2}).

\section{Diffractive Color Glass}

Nuclear targets allow to access the unitarity bound at much lower energies
than with a proton target. In fact, the central area of heavy nuclei is
"black", i.e. unitarity is saturated. As a result of saturation the
transverse momentum distribution of gluons is modified. It gets a shape
typical for the Cronin effect \cite{cronin}, i.e. gluons are suppressed at
small, but enhanced at medium-large $p_T$. This effect is called color
glass condensate (CGC) \cite{mv}.

In the nuclear rest frame the same effect looks differently, as a color
filtering of a glue-glue dipole \cite{al}. Nuclear medium resolves dipoles
of smaller size than in the case of a proton target. This results in
increased transverse momenta of radiated gluons.

First theoreticl observation of the CGC effect was made in diffraction in 
\cite{bbgg}. In a large rapidity gap process a dipole (e.g. a pion)
propagates through the nucleus experiencing color filtering as is 
illustrated in Fig.~16.
 \begin{figure}[htbp]
\includegraphics[width=8cm]{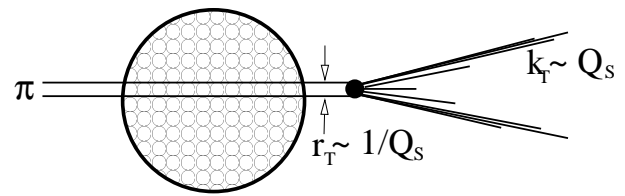}
\caption{Dipoles propagating through the nucleus experience color 
filtering leading to production of a di-jet with enhanced transverse 
momentum}
 \end{figure}
 The nuclear matter is more transparent for small size dipoles having
larger intrinsic momenta. The mean transverse momenta of quarks/jets rise
$\propto R_A$. This is a direct measurement of the saturation scale which
is expected to be
 \beqn
Q_s^2\approx 0.1\GeV^2\,A^{1/3}\approx 0.6\GeV^2
\ \ \ \ \ \ {\rm(for\ heavy\ nuclei)}
\label{2900}
 \eeqn
 This momentum is substantially larger than on a proton target.
For gluon jets the saturation scale $Q_s^2$ should be
doubled.

One may expect observation of real mini-jets at LHC.

\section{CGC and Gluon Shadowing}

The CGC leads to a rearrangement of transverse momenta of gluons 
keeping their number unaltered \cite{mv}. However, interacting gluons not 
only push each other to higher transverse momenta, but also fuse 
resulting in reduction of gluon density.
The latter effect is called shadowing. Both CGC and shadowing have the 
same origin: longitudinal overlap of gluon clouds originated from 
different bound nucleons. This is illustrated in Fig.~17.
 \begin{figure}[htbp]
\includegraphics[width=4cm]{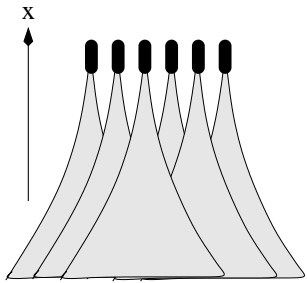}
\caption{Nucleons well separated in the longitudinal direction
in the infinite momentum frame of the nucleus create gluonic fluctuations 
which overlap at small $x$ }
 \end{figure}
 Bound nucleons in the nucleus do not overlap much, either in the rest
frame, or in the infinite momentum frame, since both the nucleon size and
internucleon spacing are subject to Lorentz contraction. However, gluons
carrying a small fraction $x$ of the proton momentum have a smaller
gamma-factor and are less compressed in the longitudinal direction.
Fig.~17 shows how gluonic clouds overlap at small $x$.

However, longitudinal overlap is not sufficient for gluon interaction, 
they must also overlap transversely. This may be a problem, since the 
transverse size of gluonic clouds is small. The mean number of overlapped 
clouds is,
 \beq
\la n_g\ra \approx \pi r_0^2\,\rho_A\,R_A\approx 0.3\ .
\label{3000}
 \eeq
 In this estimate we used the nuclear density $\rho_A=0.16\fm^{-3}$, and
the nuclear radius $R_A=7\fm$.

Thus, according to (\ref{3000}) gluons have a rather small chance to 
overlap in impact parameters even in a nucleus as heavy as lead.
Such a weak interaction of gluons leads to a weak gluon shadowing.
Fig.~18 shows the results of calculations \cite{kst2} for gluon shadowing 
in lead. 
 \begin{figure}[htbp]
\includegraphics[width=7cm]{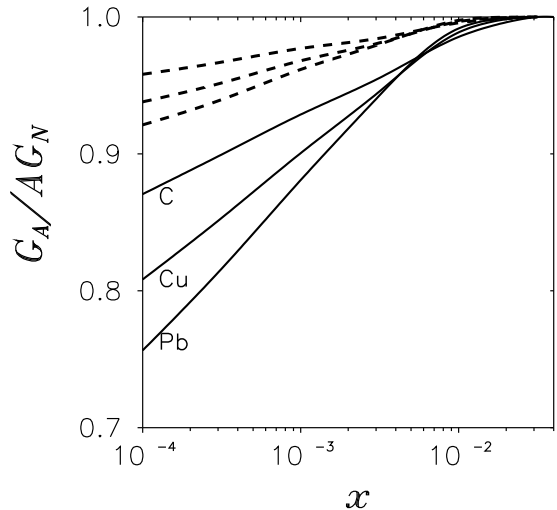}
\caption{Gluon shadowing, $G_A/G_N$ as function of Bjorken $x$ for 
carbon, copper and lead, at $Q^2=4\GeV^2$ (solid curves) and 
$Q^2=40\GeV^2$ (dashed).}
 \end{figure}
 The expected reduction of gluon density is less than $20\%$ even at very 
small $x$. This was confirmed recently in \cite{florian} by an NLO 
analysis of NMC data for DIS on nuclei.

We conclude that smallness of gluonic spots in the proton prevents them
from overlap in impact parameters even in heavy nuclei. This fact leads to
substantial reduction of gluon shadowing and of the CGC effect compared to
wide spread expectations.

\section{Summary}

\begin{itemize}

\item Fock hadronic components which are eigenstates of interaction
gain new weights in elastic scattering. The new composition formed by
the interaction with the target can be projected to new states, thus
diffractive excitation becomes possible.
 
\item
In QCD the eigenstates of the diffractive amplitude are color dipoles
which preserve their size during interaction. Color transparency is  
the major effect governing diffraction.

\item The observed puzzling smallness of high-mass diffraction related to
diffractive gluon bremsstrahlung is a direct witness for the smallness of
gluonic spots in hadrons. As a consequence, color transparency suppresses
gluon radiation and gluon shadowing.

\item Data on elastic scattering demonstrate an onset of the unitarity
bound which causes strong breakdown of Regge factorization and suppresses
diffraction. Indeed, a dramatic deviation from the Regge factorization has
been observed in data. Asymptotically the fraction of diffraction vanishes
as $1/\ln(s)$.

\item
The observed shrinkage of diffractive cone in elastic $pp$ scattering
originates mainly from onset of the unitarity bound, rather than from 
Gribov's diffusion of gluons. Data on $J/\Psi$ photoproduction confirm 
this demonstrating a weak shrinkage (in agreement with the predicted 
magnitude)

\item Color filtering leads to a dramatic increase of transverse momenta
of jets diffractively produced on nuclei. This would be a direct
measurement of the saturation scale for the diffractive color glass
condensate.

 \end{itemize}

 \begin{acknowledgments}

One of us (B.K.) is grateful to Beatriz Gay Ducati for the kind invitation
to I LAWHEP and many fruitful discussions. This work was supported in part
by Fondecyt (Chile) grants 1030355, 1050519 and 1050589, and by DFG
(Germany) grant PI182/3-1.

\end{acknowledgments}

\end{document}